# Dolomite Mineral-Inspired Equilateral Triangular-Lattice Magnets for Quantum Magnetism

Yang Zhao[1,3], Zhaoyi Li[1,2,3], Zhibin Qiu[2,1], Jianqiao Wang[2,1], Bo Wen[3], Shu Guo[2,1,*]

[1]International Quantum Academy, Shenzhen 518048, China

[2]Shenzhen Institute for Quantum Science and Engineering, Southern University of Science and Technology, Shenzhen 518055, China

[3]School of Physics and Electronics, Henan University, Kaifeng 475004, China

[*]Correspondence should be addressed to Shu. Guo (guos@sustech.edu.cn/shu-guo@outlook.com).

**ABSTRACT:** Equilateral triangular lattice magnets provide a versatile materials platform for exploring exotic quantum spin phenomena, while their field-tunable magnetic entropy offers opportunities for low-temperature adiabatic demagnetization refrigeration. Inspired by the natural mineral, we proposed a chemical strategy to achieve equilateral TL magnets, leveraging the high crystal symmetry of a large family of dolomite-type materials. As typical examples, the dolomite-type materials $SnM(BO_3)_2$ ($M$ = Co, Mn) were synthesized, and structural analysis reveals that $Co^{2+}$ and $Mn^{2+}$ ions form equilateral triangular lattices with an A-B-C stacking fashion. The magnetic susceptibilities and specific heat measurements reveal dominant antiferromagnetic interactions, with Néel temperatures of 0.49 K for $SnCo(BO_3)_2$ and 0.96 K for $SnMn(BO_3)_2$, respectively. Our results establish the dolomite-type $M'M(X)_2$ ($M'$ and $M$ sites allow various valence states, e.g., +4/+2 or +3/+3; $X$ = $CO_3^{2-}$ or $BO_3^{3-}$) system as a chemically flexible and structurally perfect material platform for exploring frustrated magnetism and low-temperature magnetocaloric applications.

Geometrically magnetic frustration stems primarily from geometric lattice constraints and competing exchange interactions.[1] In geometrically frustrated systems, incompatible antiferromagnetic (AFM) interactions enhance quantum fluctuations, which suppress conventional magnetic order and favor the emergence of exotic entangled quantum phases, including spin-nematic phases[2, 3], quantum critical behavior[4], and quantum spin liquids (QSLs)[5-7]. Geometrically frustrated lattices can typically be categorized as one-dimensional (e.g., triangular chain[8]), two-dimensional (e.g., triangular[9], Shastry-Sutherland[10, 11]), and three-dimensional (e.g., pyrochlore[12-14], garnet[15]). Among them, the triangular lattices (TLs) have attracted widespread attention as one of the simplest geometrically frustrated systems, since Philip Warren Anderson proposed QSLs in 1973[16]. Recently, equilateral TL materials have gained considerable interest owing to their ability to host exotic quantum spin states and provide promising candidates for low-temperature magnetic refrigeration.[17-21]

Despite significant progress in recent years in TL-frustrated systems, realizing equilateral TL antiferromagnets remains challenging due to material scarcity, structural disorder, and distortion. Rather than relying exclusively on the serendipitous discovery of unprecedented crystal structures, mineral-inspired materials design offers a rational strategy for identifying new geometrically frustrated magnets. The vast structural and compositional diversity of natural minerals provides a rich library of crystallographic prototypes from which ideal equilateral TL frameworks and their chemical derivatives can be systematically developed. As a typical mineral with three-fold rotation crystal symmetry, dolomite $CaMg(CO_3)_2$[22], was first discovered by Déodat Gratet de Dolomieu in the dolomite Mountains of northeastern Italy in 1791[23]. In 1934, nordenskiöldine, $SnCa(BO_3)_2$, was confirmed to be isostructural with dolomite.[24] The dolomite-type and the related calcite-type materials could be summarized with the general formula $M'M(X)_2$ ($M'$ and $M$ = metal ions with oxidation state of +1 to +5; $X$ = bridging units, $CO_3^{2-}$ or $BO_3^{3-}$), providing a versatile materials platform for investigating TL magnets. The currently known members of $M'M(X)_2$ family and their structural features are summarized in **Table 1**. Benefitting from the 3-fold rotation crystal symmetry, both the $M'$ and $M$ cations sit on the equilateral TLs. The $M'$ and $M$ sites can accommodate various valence-state combinations, such as +5/+1, +4/+2, +3/+3, and +2/+2, while $X$ represents bridging units ($CO_3^{2-}$ or $BO_3^{3-}$). These compounds have garnered significant attention for their unique properties in thermal stability[25], magnetism[26], and luminescence[27]. To the best of our knowledge, only a limited number of members have been successfully synthesized in this family[28-31], and even fewer studies have examined their magnetic properties, leaving the magnetism of this family largely unexplored.

Inspired by the rich diversity of compounds within the $M'M(X)_2$ family and the limited understanding of their magnetic properties, this work investigates the magnetic and thermodynamic behaviors of the dolomite-structured TL materials $SnM(BO_3)_2$ ($M$ = Co, Mn). Although the structures of $SnM(BO_3)_2$ ($M$ = Co, Mn) have been reported[32], their magnetic properties, particularly the low-temperature magnetic states arising from geometric frustration, remain largely unexplored. The polycrystalline samples of $SnM(BO_3)_2$ ($M$ = Co, Mn) were synthesized via a standardized high-temperature solid-state method (see details in **Supporting Information and Figure S1**) and subsequently used for magnetic studies. As schemat-ically shown in **Figure 1a**, different from the Calcite-type structure, $SnCo(BO_3)_2$ (SCBO) and $SnMn(BO_3)_2$ (SMBO) tend to form a relatively ordered structure with *M'* and *M* cations site into two distinct crystallographic sites. Consequently, alternating *M'* and *M* layers are stacked along the *c* axis. The $SnM(BO_3)_2$ (*M* = Co, Mn) crystallizes in a centrosymmetric rhombohedral structure with space group $R\overline{3}$ (No. 148). As shown in **Figure 1b**, the unit cell consists of layered units formed by $[MO_6]$ (*M* = Co, Mn) and $[SnO_6]$ octahedra, which are interconnected via planar triangular $[BO_3]$ groups. In this structure, $Sn^{2+}$ ions occupy the *M'* sites, whereas magnetic $Co^{2+}/Mn^{2+}$ ions occupy the *M* sites, giving rise to magnetic TL layers that are well separated by nonmagnetic $Sn^{2+}$ layers. Within the magnetic TL (**Figure 1c**), the $[MO_6]$ (*M* = Co, Mn) octahedra are connected via two alternating types of triangular $[BO_3]$ groups, generating a *M*–O–B–O–*M* (*M* = Co, Mn) super-super-exchange magnetic interaction path with nearest-neighbor (NN) distance $d_{NN}$ =4.72-4.77 Å. Although the overall structure is identical, the larger ionic radius of $Mn^{2+}$ relative to $Co^{2+}$ results in an expansion of the magnetic layers in SMBO in the main crystallographic directions. Consequently, both the in-plane NN distance ($d_{NN}$) and the interlayer separation ($d_{inter}$) are larger in SMBO (4.77 and 5.78 Å, respectively) than in SCBO (4.72 and 5.67 Å, respectively). Along the *c*-axis, the magnetic TLs are stacked in an A-B-C stacking fashion (**Figure 1d**). Due to the A-B-C-type stacking of the magnetic TL planes in this rhombohedral crystal structure, the magnetic framework forms pyramidal units composed of in-plane TLs and $d_{NN}$ magnetic ions from adjacent layers, implying potential interlayer geometric frustration[33]. The related Rietveld refinement results were summarized in **Figure S2** and **Table S1**.

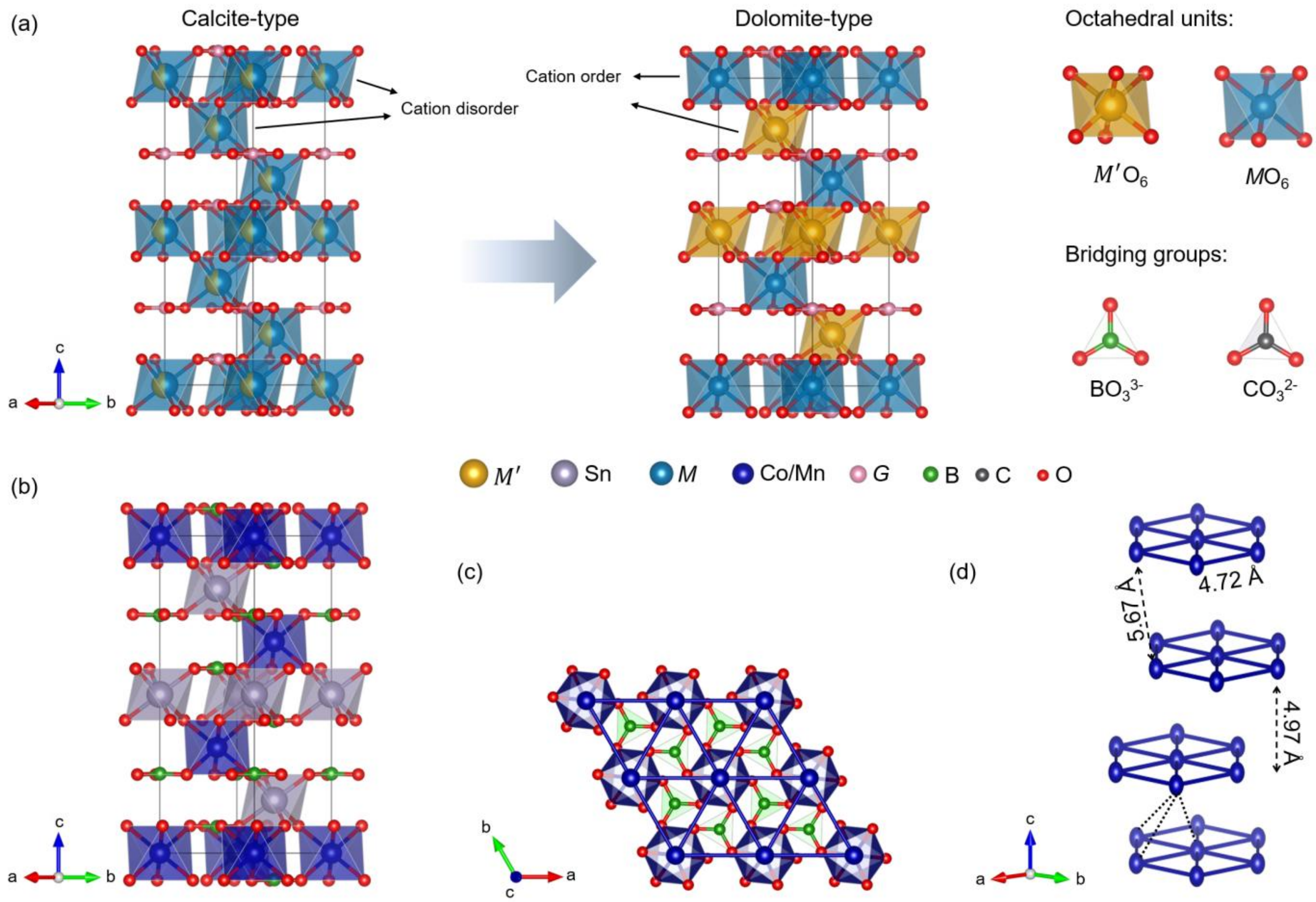

**Figure 1. Structure evolution and crystal structure.** (a) Schematic structural evolution from calcite-type (left) to dolomite-type (right) crystal structures with the general formula $M'M(X)_2$. (b) Unit cell of $SnM(BO_3)_2$ (*M* = Co, Mn). (c) Projection of the TL on the *a*-*b* plane. (d) Magnetic frameworks of $SnM(BO_3)_2$ (*M* = Co, Mn), with SCBO selected as a representative compound for illustration. G represents the general elements for triangular $GO_3$ units.

To further investigate the magnetic properties of $SnM(BO_3)_2$ (*M* = Co, Mn), direct current (DC) magnetic susceptibility $\chi(T)$ was measured. No zero-field-cooled (ZFC) and field-cooled (FC) divergence is observed in the temperature range of 1.8–50 K at 0.05 T (**Figures 2a and 2b**). Under an applied field of 0.2 T, $\chi(T)$ increases continuously upon cooling from 300 to 2 K, with no obvious anomaly indicative of long-range magnetic ordering (LRMO) above 1.8 K for SCBO and SMBO (**Figures 2c and 2d**). In the case of SCBO, a sharp anomaly in $\chi(T)$ around 0.61 K signals the onset of LRMO (inset of **Figure 2c**). Additionally, measured in the 0.2–14 T range (**Figure S3**), SCBO shows typical paramagnetic behavior throughout, and also the magnetization diminishes with field, reflecting the tendency of paramagnets to saturate at high fields. The $\chi^{-1}(T)$ for the above two materials was fitted using the Curie-Weiss (C-W) law, $\chi = \frac{C}{T-\Theta}$, where Θ is the Weiss temperature, and *C* is the Curie constant[34]. The obtained Weiss temperatures, from fits in the ranges 2–20 K and 50–200 K for SCBO and SMBO, respectively, are $\Theta_{Co}$ = -1.45 K and $\Theta_{Mn}$ = -6.62 K, indicating dominant antiferromagnetic (AFM) interactions. The effective magnetic moments $\mu_{eff}$, calculated via $\mu_{eff} = \sqrt{8C}\,\mu_B$, are 4.16 $\mu_B$ for SCBO and 6.16 $\mu_B$ for SMBO (**Table S2**). The $\mu_{eff}$ of SMBO agrees well with the theoretical spin-only value of ~5.92 $\mu_B$ for high-spin $Mn^{2+}$ ($S$ = 5/2), whereas that of SCBO deviates from the spin-only value of ~1.73 $\mu_B$ for low-spin $Co^{2+}$ ($S_{eff}$ = 1/2, $g$ = 2), which may be related to the strong magnetic anisotropy of the $Co^{2+}$ ions [35] and the possible preferred orientation of the powder sample during the pressing process [36].

The isothermal magnetization curves $M(B)$ of SCBO and SMBO were measured under fields up to 14 T (**Figures 2e and 2f**). For SCBO, the nearly linear increase above 3 T might be attributed to Van Vleck paramagnetism arising from excited-state mixing. After subtracting this background, the magnetization saturates at about 2.13 $\mu_B$ (**Figure S4**). The experimentally observed saturation moment exceeds the spin-only value of 1 $\mu_B$ expected for an ideal effective $S_{eff}$ = 1/2 state with $g$ = 2, which may arise from an enhanced effective $g$ factor associated with the spin-orbit coupling and the partially unquenched orbital angular momentum of $Co^{2+}$. At 0.4 K, the derivative $\partial M/\partial B$ of the $M(B)$ curve for SCBO reaches its maximum at 0.4 T, indicating a strong magnetic response associated with the spin-alignment process (**Figure S5**). For SMBO, the magnetization saturates above 7 T at approximately 5.02 $\mu_B$ per $Mn^{2+}$, which closely matches the free $Mn^{2+}$ ion (5 $\mu_B$ for $S$ = 5/2, $L$ = 0), indicating full spin polarization in the high-field state. Further analysis of the magnetic response for SMBO at low temperatures was conducted by taking the derivative of the magnetization curve at 1.8 K (**Figure 2f**). The peak $\partial M/\partial B$ value in the low-field region indicates a strong magnetic response. As the field increases, $\partial M/\partial B$ approaches zero near 8 T for SMBO, indicating the onset of magnetic saturation.

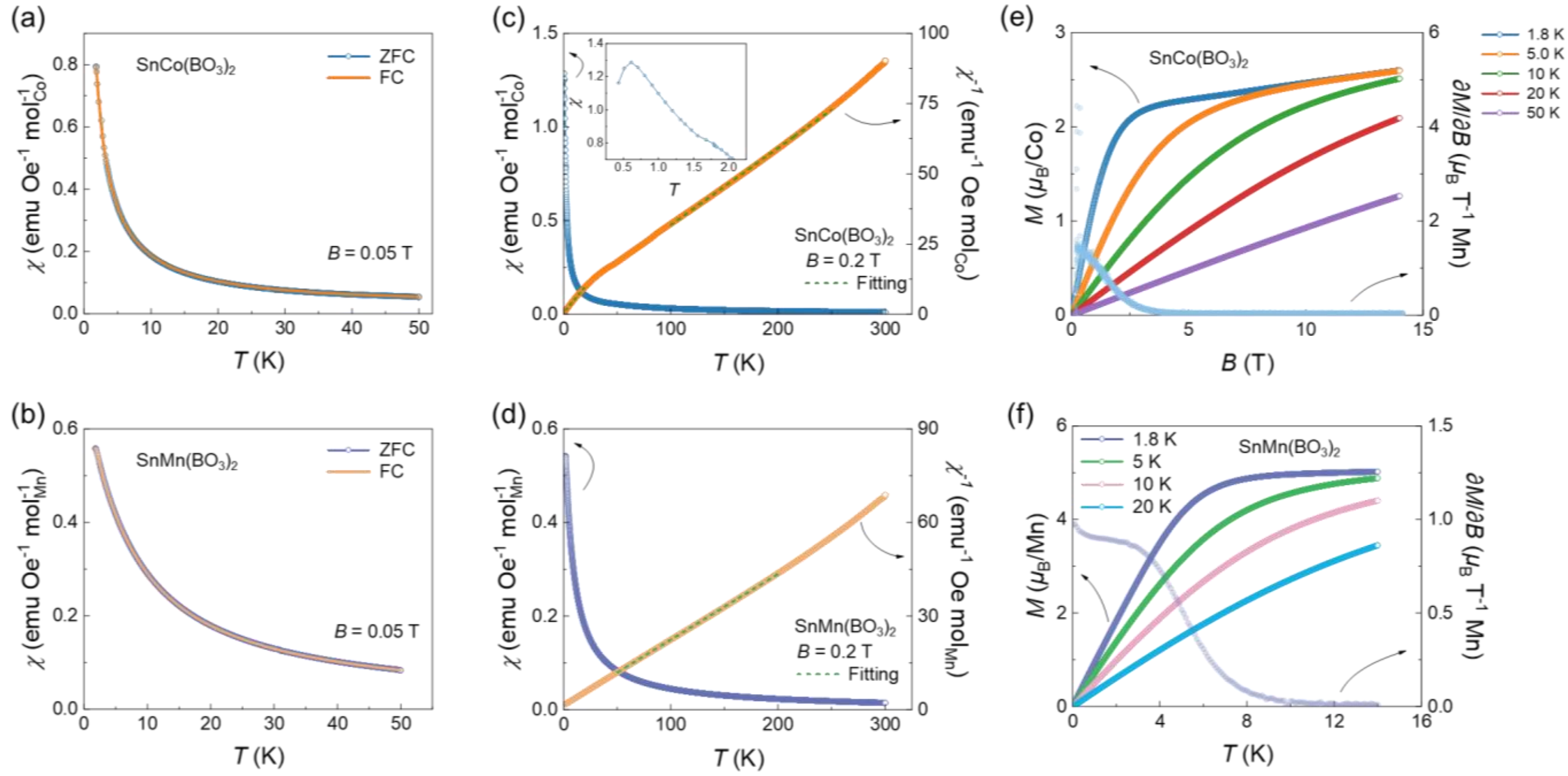

**Figure 2 Magnetism**. The figures respectively show the magnetic measurement results of SCBO and SMBO magnets: (a, b) ZFC and FC DC magnetic susceptibility $\chi(T)$ curves measured under a magnetic field of 0.05 T; (c, d) DC $\chi(T)$ and its inverse $\chi^{1}(T)$ curves obtained at 0.2 T, including C-W fitting; the insets show enlarged $\chi(T)$ data collected between 1.8 and 0.4 K using the $^{3}$He option; (e, f) isothermal magnetization curves $M(B)$ over the magnetic field range of 0–14 T. Light blue and purple solid circles: $\partial M/\partial B$ for SCBO and SMBO at 1.8 K, respectively.

To further examine the low-temperature behavior, we performed specific heat measurements at temperatures above 0.4 K (**Figures 3a and 3b, and Figure S6**). In zero field, the specific heat of both magnets exhibits sharp $\lambda$-shaped peaks, marking the LRMO from the paramagnetic to the AFM ordering, yielding Néel temperatures $T_N$ = 0.49 K for SCBO and $T_N$ = 0.96 K for SMBO, respectively. Benefitting from the low transition temperatures and relatively high magnetic ion densities, SCBO and SMBO are favorable for adiabatic demagnetization refrigeration (ADR) in the sub-Kelvin region. Further investigation reveals that the $\lambda$-peaks of both SCBO and SMBO progressively shift to lower temperatures as the applied field increases, characteristic of AFM transitions. The magnetic ordering temperatures of $SnM(BO_3)_2$ ($M$ = Co, Mn) are significantly suppressed to values well below their corresponding C-W temperatures, with the magnetic frustration parameters $f = |\Theta|/T_N$ reaching 2.96 and 6.90, respectively (**Table S2**).

To isolate the magnetic specific heat, $C_M$, from the original specific heat curves, we employed a two-Debye model[37,38] to account for the phonon contribution ($C_{p,ph}$). By introducing two Debye temperatures, this model provides a more accurate description of the magnet's heat capacity behavior over a broad temperature range: $C_{p,ph} = 9N_A k_B \sum_{n=1}^{2} C_n \left(\frac{T}{\Theta_{Dn}}\right)^3 \int_0^{\frac{\Theta_{Dn}}{T}} \frac{x^4 e^4}{(e^x-1)^2} dx$, where $N_A$ is Avogadro's constant, $k_B$ is Boltzmann's constant, and the weighting coefficients satisfy $C_1 + C_2$ = atomicity. The Two-Debye fits, shown as black dashed lines in **Figures 3c and 3d**, agree well with the experimental data. The phonon contribution to the specific heat can be obtained by extrapolating the fitted curves, shown as solid lines in **Figures 3c and 3d**. Considering the insulating features of $SnM(BO_3)_2$ ($M$ = Co, Mn), the electronic contribution can be neglected. The resulting zero-field $C_M/T$ is plotted in **Figures 3e and 3f**. The magnetic entropy, $S_M$, was calculated by integrating $C_M/T$ over temperature: $S_M = \int_0^T (C_M/T) dT$. The recovered magnetic entropies of $S_M$ of SCBO and SMBO reach 4.63 and 13.55 J mol$^{-1}$ K$^{-1}$, respectively, which are close to the theoretical values expected for $S_{eff}$ = 1/2 (Rln2 ≈ 5.76 J mol$^{-1}$ K$^{-1}$) and $S$ = 5/2 (Rln6 ≈ 14.90 J mol$^{-1}$ K$^{-1}$). The remaining magnetic entropy might have originated from the finite lower temperature limits of the heat-capacity measurements. $Co^{2+}$ often exhibits a pseudospin $S_{eff}$ = 1/2 state at low temperatures, a behavior similarly observed in compounds such as $Na_2BaCo(PO_4)_2$ [18, 39, 40]. And the magnetic entropy $S_M$ released near the phase transition is only 0.21Rln2 for SCBO and 0.38Rln6 for SMBO, suggesting the residual spin fluctuations may persist into the low-temperature ordered state. To further assess their potential for ADR, the maximum magnetic entropy change ($-\Delta S_M^{max}$) was evaluated. The substitution of $Co^{2+}$ by $Mn^{2+}$ increases the $-\Delta S_M^{max}$ from 13.71 to 21.43 J kg$^{-1}$ K$^{-1}$ (**Figure S7**), suggesting potential sub-kelvin ADR applications.

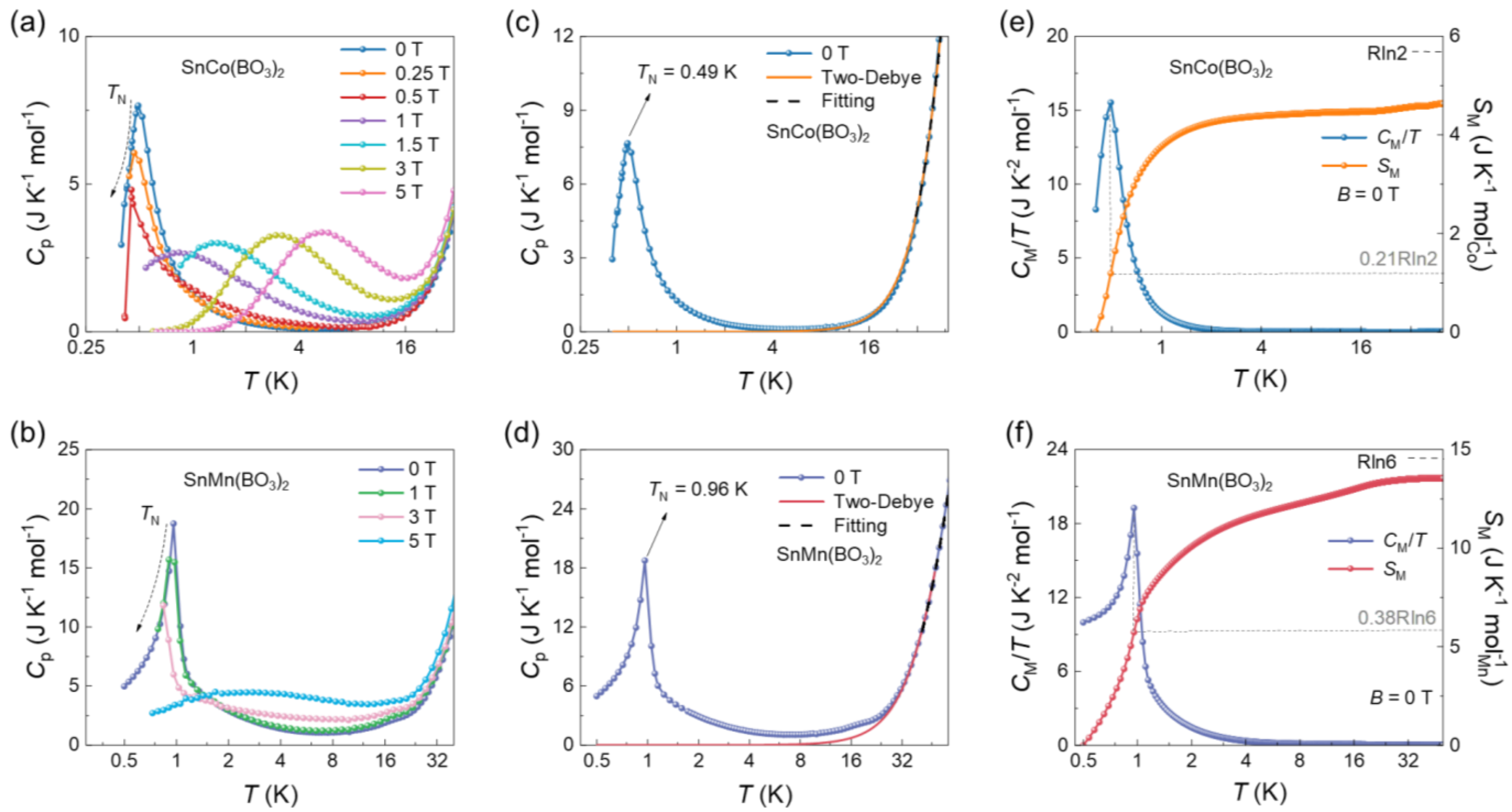

**Figure 3. Thermodynamics.** (a, b) Temperature-dependent specific heat $C_p(T)$ of $SnM(BO_3)_2$ ($M$ = Co, Mn) under various applied magnetic fields. (c, d) Zero-field $C_p(T)$, where the black dashed lines indicate the phonon contribution obtained from fitting the zero-field $C_p(T)$ with a Two-Debye model, and the solid lines correspond to extrapolations of these fits. (e,f) Zero-field magnetic specific heat divided by temperature, $C_M/T$, and the associated magnetic entropy $S_M$.

**Table 1. Crystal chemistry of the $M'M(X)_2$ family: structure type, valence-state combination, and known compound.**

| Structure type | Dolomite | Dolomite | Calcite | Dolomite | Dolomite |
|---|---|---|---|---|---|
| Space group | $R\bar{3}$ | $R\bar{3}$ | $R\bar{3}c$ | $R\bar{3}$ | $R\bar{3}$ |
| Valence-state combination ($M'$, $M$) | (+4, +2) | (+3, +3) | | (+2, +2) | (+1, +5) |
| Compounds | $SnMg(BO_3)_2$ [32]<br>$Sn\mathbf{Ni}(BO_3)_2$ [32]<br>$Sn\mathbf{Co}(BO_3)_2$ [32]<br>$Sn\mathbf{Mn}(BO_3)_2$ [32]<br>$Sn\mathbf{Fe}(BO_3)_2$ [28]<br>$SnCa(BO_3)_2$ [24]<br>$SnCd(BO_3)_2$ [32]<br>$SnSr(BO_3)_2$ [32]<br>$SnBa(BO_3)_2$ [32]<br>$ZrCa(BO_3)_2$ [25]<br>$ZrSr(BO_3)_2$ [25]<br>$ZrBa(BO_3)_2$ [25]<br>$ZrCd(BO_3)_2$ [25]<br>$TiBa(BO_3)_2$ [25]<br>$PbCa(BO_3)_2$ [29] | $\mathbf{Cr}Lu(BO_3)_2$ [30]<br>$\mathbf{CrYb}(BO_3)_2$ [30]<br>$\mathbf{CrEr}(BO_3)_2$ [30]<br>$\mathbf{Cr}Y(BO_3)_2$ [30]<br>$\mathbf{CrDy}(BO_3)_2$ [30]<br>$\mathbf{CrHo}(BO_3)_2$ [26]<br>$\mathbf{CrTm}(BO_3)_2$ [26] | $ScY(BO_3)_2$ [26]<br>$Sc\mathbf{Ho}(BO_3)_2$ [26]<br>$Sc\mathbf{Er}(BO_3)_2$ [26]<br>$Sc\mathbf{Tm}(BO_3)_2$ [26]<br>$Sc\mathbf{Yb}(BO_3)_2$ [26]<br>$ScLu(BO_3)_2$ [26]<br>$In\mathbf{Cr}(BO_3)_2$ [30]<br>$InLu(BO_3)_2$ [30] | $CaMg(CO_3)_2$ [22]<br>$Ca\mathbf{Fe}(CO_3)_2$ [41]<br>$Ca\mathbf{Mn}(CO_3)_2$ [41] | $NaNb(BO_3)_2$ [29] |

Note: magnetic cations are shown in bold.

To facilitate the comparison, previously reported dolomite-type and related calcite-type compounds are summarized in **Table 1.** These two types of materials share the same chemical formula $M'M(X)_2$. Unlike site-mixing in calcite-type materials, the $M'$ and $M$ sites in the crystal structure of dolomite-type materials are relatively ordered, which is favorable for constructing geometrically frustrated magnetic lattices with minimized structural disorder. More importantly, $M'$ and $M$ sites can accommodate various valence combi-nations (+2/+2, +4/+2, +3/+3, +1/+5) and different cationic sizes, expanding the possibilities of different quantum spin numbers and spatial arrangements. On the one hand, small quantum spin number magnetic systems usually exhibit relatively strong quantum fluctuations and are favorable for investigating novel quantum spin states. On the other hand, TL systems with large spin numbers usually cover more classic magnetic behaviors and exhibit greater magnetic entropy, making them suitable for low-temperature magnetocaloric applications. Furthermore, as bridging units in the TL layers, $BO_3^{3-}$ and $CO_3^{2-}$ could enable different intralayer exchange pathways with magnetic interactions.

In summary, this study investigates the magnetic and thermodynamic properties of the dolomite-type compounds $SnM(BO_3)_2$ ($M$ = Co, Mn). Magnetic susceptibility and specific heat measurements revealed AFM orderings at approximately 0.49 and 0.96 K in SCBO and SMBO, respectively. More importantly, this work highlights the broader potential of the dolomite-type $M'M(X)_2$ family as a highly versatile platform for realizing the geometrical frustrated TL materials. Future studies involving alternative magnetic ions, bridge-group substitution, cation valence engineering, and partial cation substitution are expected to further expand the accessible magnetic phase space and facilitate the discovery of novel frustrated magnets, quantum states, and multifunctional magnetic materials.

## AUTHOR INFORMATION

### Corresponding Author

**Shu Guo** – Shenzhen Institute for Quantum Science and Engineering, Southern University of Science and Technology, Shenzhen 518055, China; International Quantum Academy, Shenzhen 518048, China; orcid.org/0000-0002-20988904; Email: guos@sustech.edu.cn/shuguo@outlook.com.

### Authors

**Yang Zhao** – International Quantum Academy, Shenzhen 518048, China; School of Physics and Electronics, Henan University, Kaifeng 475004, China.

**Zhaoyi Li** – Shenzhen Institute for Quantum Science and Engineering, Southern University of Science and Technology, Shenzhen 518055, China; School of Physics and Electronics, Henan University, Kaifeng 475004, China; International Quantum Academy, Shenzhen 518048, China.

**Zhibin Qiu** – Shenzhen Institute for Quantum Science and Engineering, Southern University of Science and Technology, Shenzhen 518055, China; International Quantum Academy, Shenzhen 518048, China.

**Jianqiao Wang** – Shenzhen Institute for Quantum Science and Engineering, Southern University of Science and Technology, Shenzhen 518055, China; International Quantum Academy, Shenzhen 518048, China.

**Bo Wen** – School of Physics and Electronics, Henan University, Kaifeng 475004, China; orcid.org/0000-0001-7252918X.

### Notes

The authors declare no competing interests.

## ACKNOWLEDGMENT

The authors acknowledge the financial support from the National Natural Science Foundation of China (22205091), and the Guangdong Pearl River Talent Plan (2023QN10C793).

Supporting information

# Dolomite Mineral-Inspired Equilateral Triangular-Lattice Magnets for Quantum Magnetism

## Experimental Details and Extended Discussion

### Material Synthesis

Polycrystalline samples of SCBO and SMBO were prepared using a high-temperature solid-state reaction method. High-purity starting materials, including $SnO_2$ (99.99%), CoO (99.99%), MnO (99.99%), and $H_3BO_3$ (99.998%), were mixed in slightly non-stoichiometric ratios to compensate for differences in volatility and thermal stability among them. The thoroughly ground and homogenized reagents were first calcined at 600 ℃ for 6 hours in air to decompose $H_3BO_3$ and remove moisture. Subsequently, pre-sintering was conducted for 6 hours in air at 1100℃ (SCBO) or 1050℃ (SMBO) to improve interparticle contact and prevent incomplete reactions during final sintering. Subsequently, both samples were further ground and finally sintered in air at 1100°C (SCBO) and 1050°C (SMBO) for 10 hours, yielding light purple and brown polycrystalline powders, respectively, as shown in **Figure S1**.

### Structural Analysis

The polycrystalline sample was evenly spread on a dedicated glass slide and then secured on the sample stage of a Rigaku SmartLab 9 kW X-ray diffractometer for testing to acquire P-XRD patterns. The experiment employed a Bragg-Brentano optical system, utilizing Cu $K_a$ radiation ($K_{a1}$ = 1.54050 Å, $K_{a2}$ = 1.54430 Å) as the X-ray source. Key parameter settings included a scanning range of 5° to 110°, a step size of 0.02°, and a scanning rate of 0.9° per minute. The diffraction patterns obtained from the experiment were structurally refined using the GSAS2 software system [42]. The relevant unit cell parameters are summarized in **Table S1**. The Rietveld refinements of the room-temperature powder X-ray diffraction (P-XRD) patterns (**Figures S2**) show excellent agreement between simulation and experiment. The low values of the reliability factors ($R_{wp}$ = 1.57% and $\chi^2$ = 3.14 for SCBO; $R_{wp}$ = 3.45% and $\chi^2$ = 4.02 for SMBO) confirm the accuracy of the refined structural models for $SnM(BO_3)_2$ ($M$ = Co, Mn).

### Magnetic properties

The direct current (DC) magnetic susceptibility of the SCBO and SMBO was measured using the Vibrating Sample Magnetometer (VSM) option of a Quantum Design Physical Property Measurement System (PPMS) at temperatures above 1.8 K. The powder sample masses used were 15.35 mg and 9.17 mg for SCBO and SMBO, respectively. Furthermore, in the temperature range down to 0.4 K, measurements were conducted on a 4.3 mg SCBO sample using the Superconducting Quantum Interference Device (SQUID) magnetometer option of a Magnetic Property Measurement System (MPMS3) equipped with a $^3$He insert. The specific heat of SCBO and SMBO was measured using a PPMS equipped with a $^4$He module (temperature range: 1.8–200 K) and a $^3$He module (temperature range: > 0.4 K). For measurements with the $^4$He module, polycrystalline samples of SCBO and SMBO weighed 2.13 mg and 1.79 mg, respectively (**Figure S6a and S6b**). The $^3$He module measurements employed samples of 0.88 mg (SCBO) and 0.65 mg (SMBO), which were calibrated against their respective high-temperature specific heat data (**Figure S6c and S6d**).

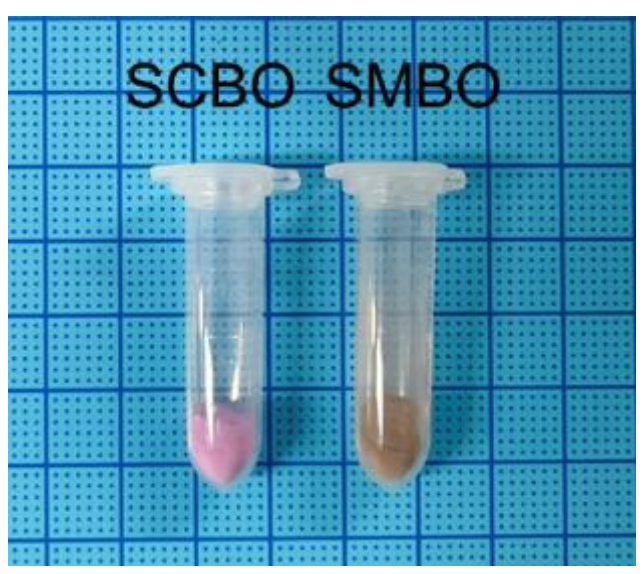

**Figure S1** The polycrystalline sample image, with $SnCo(BO_3)_2$ on the left and $SnMn(BO_3)_2$ on the right.

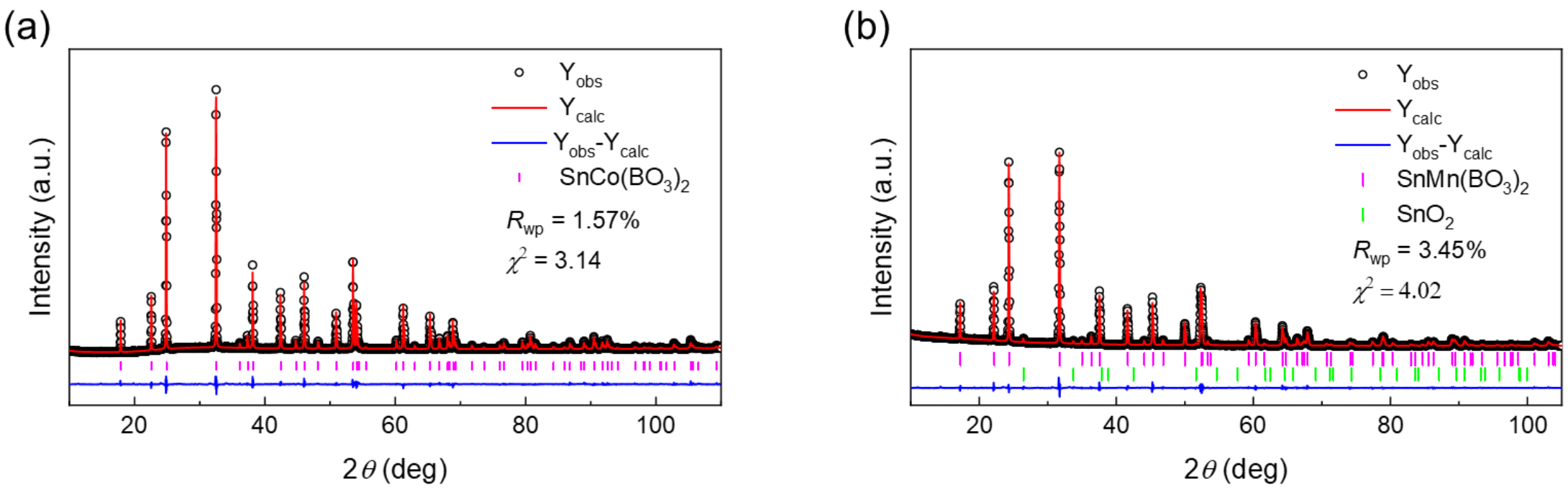

**Figure S2** (a) and (b) Rietveld P-XRD refinement for SCBO and SMBO, respectively. The observed (black circles), calculated (red solid line), and difference ($Y_{obs}$ - $Y_{calc}$, blue) X-ray patterns are presented. The positions of Bragg reflections are marked by the pink (SCBO or SMBO) and green ($SnO_2$) tick marks.

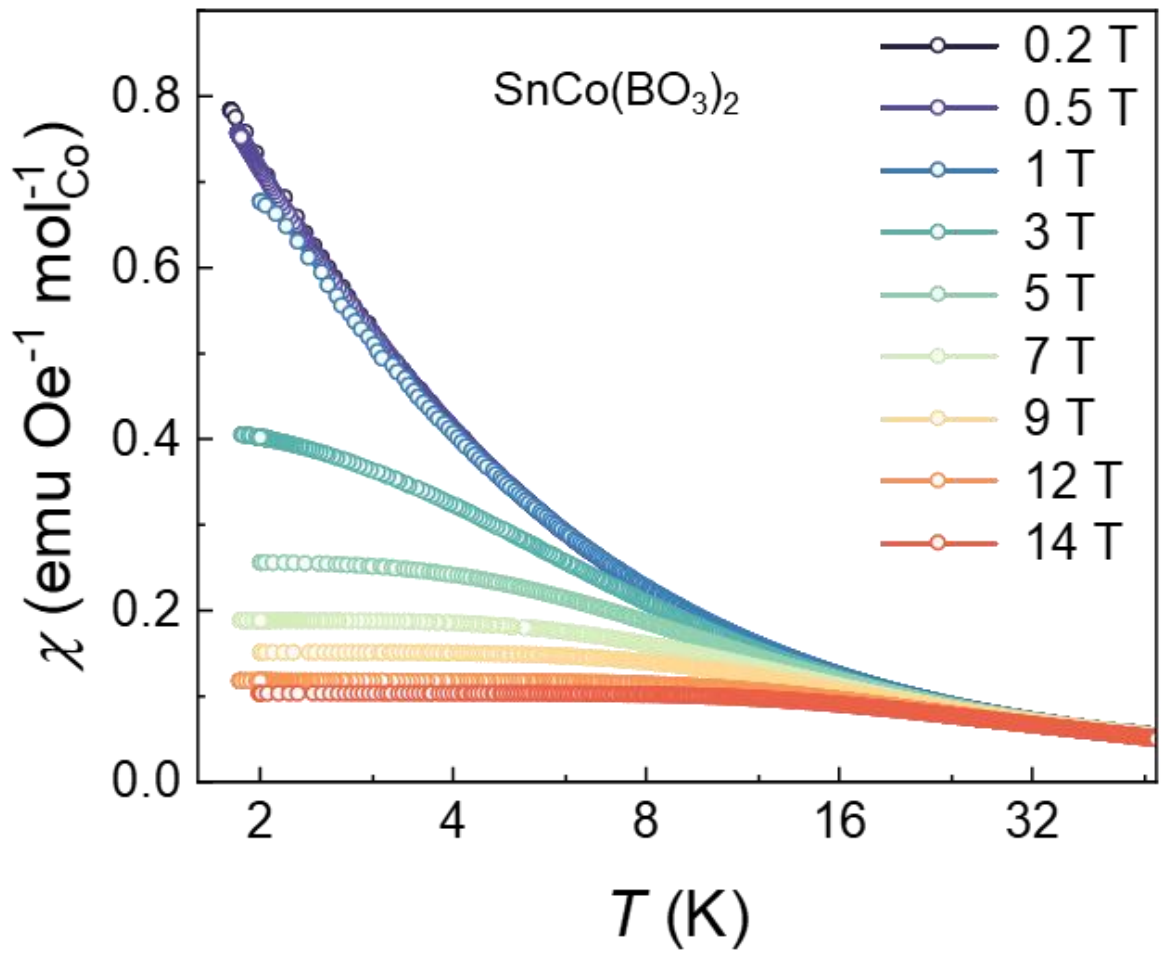

**Figure S3** The dependence of DC magnetic susceptibility on temperature in the range of 1.8–50 K under varying magnetic fields up to 14 T.

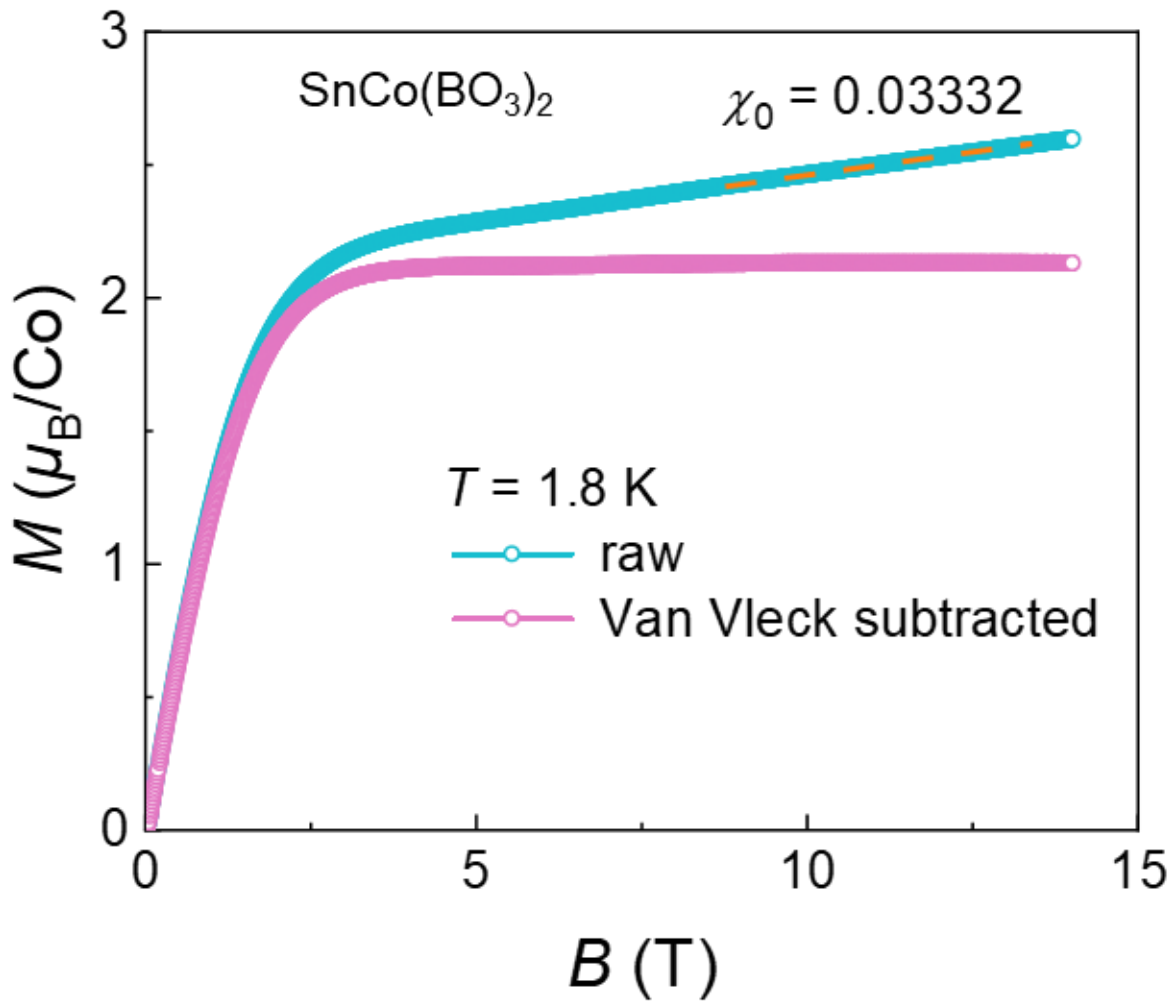

**Figure S4** Isothermal magnetization curves at 1.8 K, with Van Vleck paramagnetic contribution subtracted.

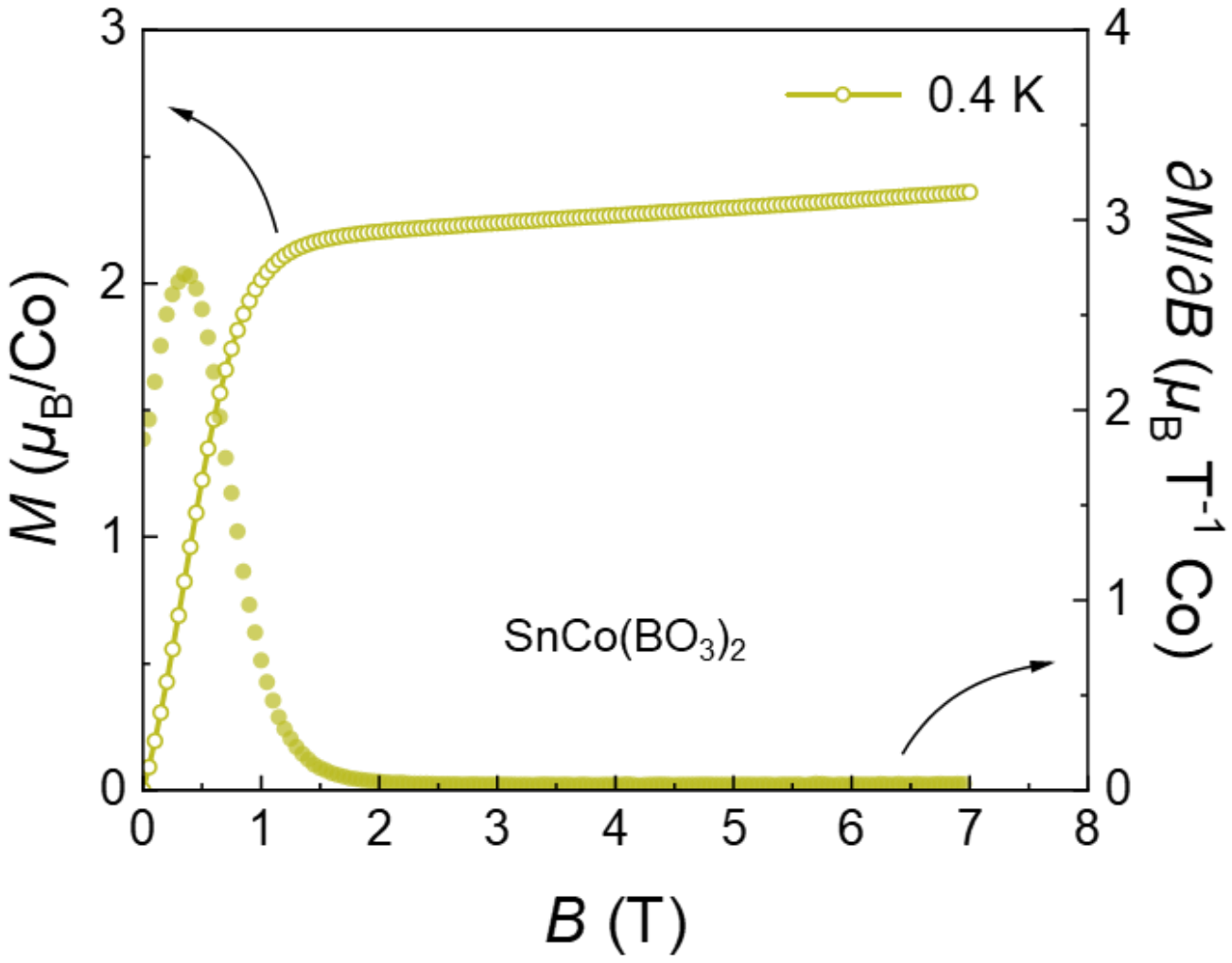

**Figure S5** The isothermal magnetization curve of SCBO at $T = 0.4$ K and its derivative $\partial M/\partial B$.

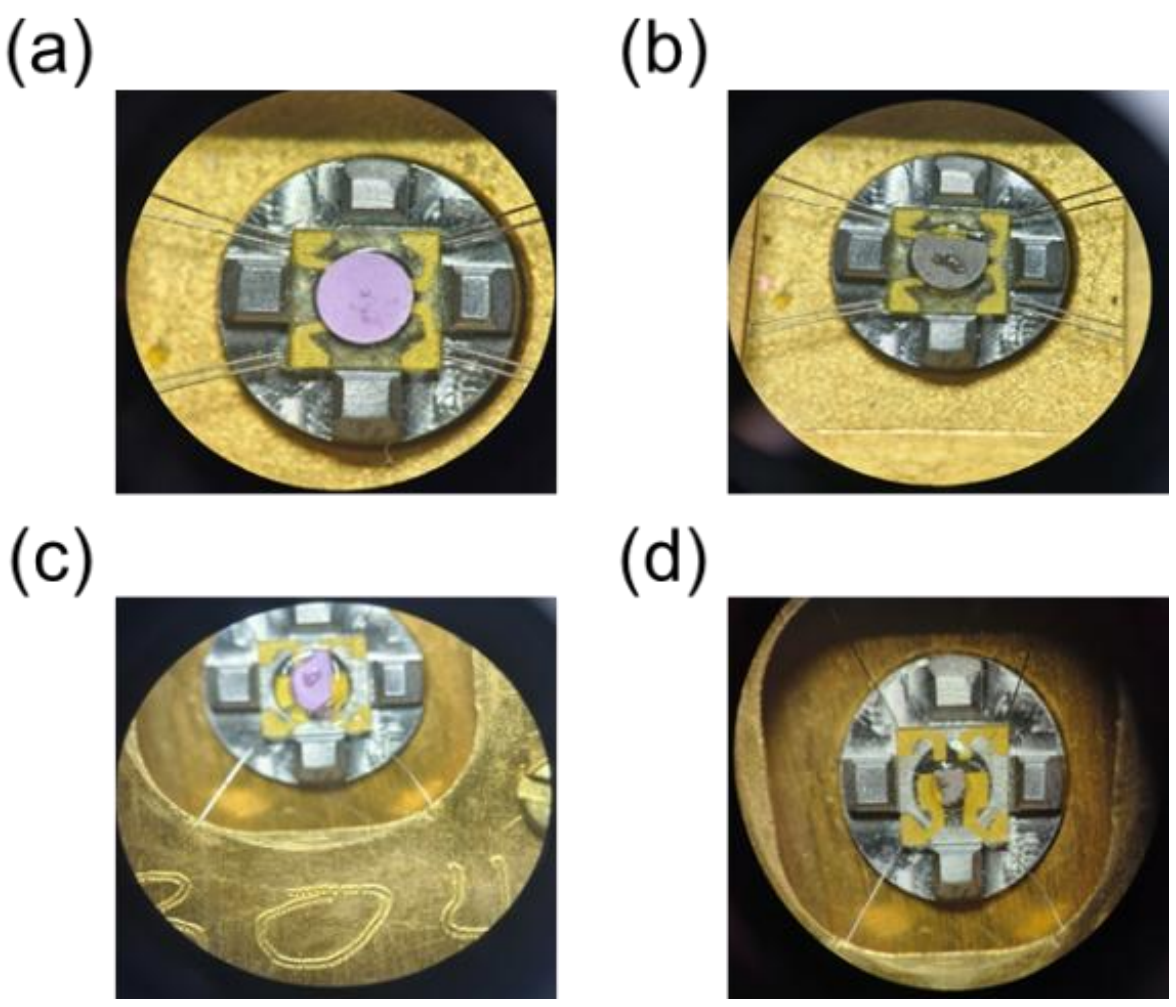

**Figure S6** (a) and (b) The $^4$He specific heat measurements for SCBO and SMBO, respectively, cooled down to 1.8 K. (c) and (d) the specific heat measurements using a $^3$He module for SCBO and SMBO, with temperatures reaching as low as 0.4 K and 0.5 K, respectively.

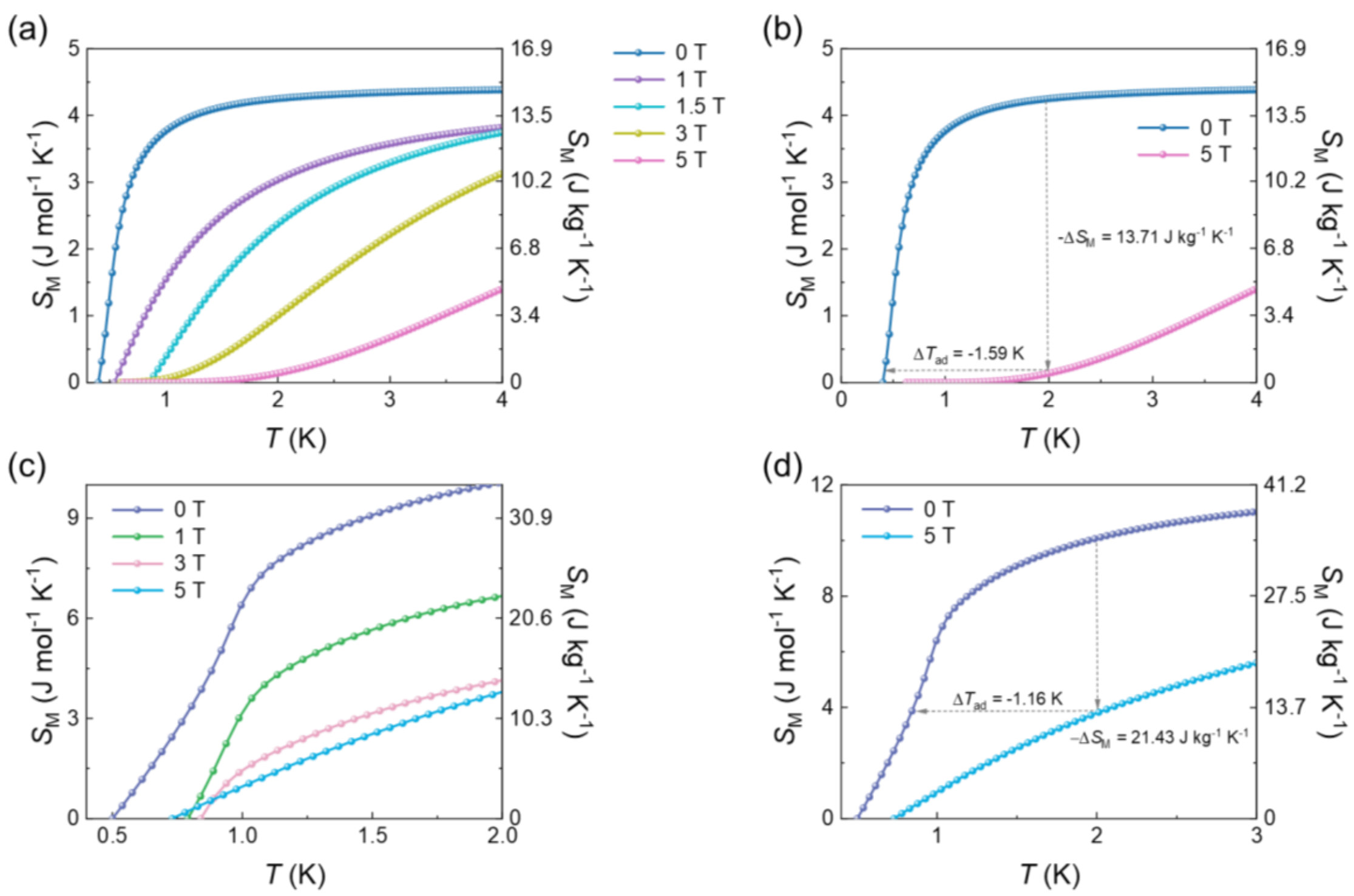

**Figure S7** (a) and (c) Temperature dependence of magnetic entropy for SCBO and SMBO under different magnetic fields, respectively; (b) and (d) Simulated isothermal magnetization and adiabatic demagnetization processes for SCBO and SMBO at an initial temperature of 2 K and an initial magnetic field of 5 T, respectively.

**Table S1.** Powder X-ray data Rietveld refined parameters of Sn*M*$(BO_3)_2$ (*M* = Co, Mn).

| Empirical formula | $SnCo(BO_3)_2$ | $SnMn(BO_3)_2$ |
|---|---|---|
| Space group | $R\bar{3}$ | $R\bar{3}$ |
| Formula weight | 878.29 | 866.53 |
| Temperature/K | 300 | 300 |
| $a$/Å | 4.72254(5) | 4.76801(7) |
| $b$/Å | 4.72254 | 4.76801 |
| $c$/Å | 14.91077(11) | 15.31251(18) |
| $\alpha$/° | 90 | 90 |
| $\beta$/° | 90 | 90 |
| $\gamma$/° | 120 | 120 |
| Cell volume /$Å^3$ | 287.993(5) | 301.475(8) |
| Z | 1 | 1 |
| $\rho_{calc}$ /$g{\cdot}cm^{-3}$ | 5.0641 | 4.7729 |
| Radiation/Å | Cu $K_{\alpha1,2}$ ($\lambda_1$ = 1.54050, $\lambda_2$ = 1.54430) | Cu $K_{\alpha1,2}$ ($\lambda_1$ = 1.54050, $\lambda_2$ = 1.54430) |
| 2θ range for data collection/° | 5.00-109.96 | 5.00-109.96 |
| Independent reflections | 80 | 108 |
| $\chi^2$ | 3.14 | 4.02 |
| $R_p$ | 0.00991 | 0.02250 |
| $R_{wp}$ | 0.01570 | 0.03447 |

**Table S2** C-W temperature Θ and effective moment $\mu_{eff}$ were fitted from different regions. Néel temperature $T_N$ was obtained from specific heat, enabling calculation of the frustration factor *f*.

| | SCBO | | SMBO |
|---|---|---|---|
| Fitting temperature range (K) | 2 – 20 | 100 – 250 | 50 – 200 |
| Θ (K) | -1.45 | -16.19 | -6.62 |
| $\mu_{eff}$ ($\mu_B$) | 4.16 | 5.41 | 6.16 |
| $T_N$ (K) | 0.49 | | 0.96 |
| $f$ | 2.96 | | 6.90 |